\documentclass[linenum]{revtex4-2}

\usepackage{graphicx}
\usepackage{xcolor}
\usepackage{soul}

\usepackage{comment}

\newcommand{\change}[1]{{\color{cyan}#1}}

\begin{document}

%\title{Measuring sea ice mechanical properties using Distributed Acoustic Sensing}

\title{High-resolution measurement of sea ice mechanical characteristics using Distributed Acoustic Sensing}

\author{Sébastien Kuchly$^1$, Ludovic Moreau$^2$, Vasco Zanchi $^1$, Nicolas Mokus$^2$, Véronique Dansereau $^{2,3}$,Madison M. Smith$^4$, Dany Dumont$^5$, Stéphane Perrard $^1$, Antonin Eddi $^1$}

%\affil[1]{First Affiliation, Institute, City, Country}
%{\auorc{0000-0000-0000-0001}{(AAF)}\auorc{0000-0000-0000-0002}{(ABS)}}
%\affil[2]{Second Affiliation, Institute, City, Country}
%{\auorc{0000-0000-0000-0003}{(ACT)}}
%\affil[3]{Third Affiliation, Institute, City, Country}
%{}
%\affil[4]{Third Affiliation, Institute, City, Country}
%{}

\affiliation{$^1$PMMH Laboratory, ESPCI, CNRS, PSL University, Sorbonne Université, Université Paris Cité, 7 Quai Saint Bernard, 75005 Paris, France}
\affiliation{$^2$ISTerre, Univ. Grenoble Alpes, Univ. Savoie Mont Blanc, CNRS, IRD, Univ. Gustave Eiffel, ISTerre, 38000 Grenoble, France}
\affiliation{$^3$IGE, Univ. Grenoble Alpes, CNRS, IRD, Grenoble INP, IGE, 38000 Grenoble, France}
\affiliation{$^4$Woods Hole Oceanographic Institution, Woods Hole, MA, USA}
\affiliation{$^5$Institut des sciences de la mer, Universit\'e du Qu\'ebec \`a Rimouski, Rimouski, Qu\'ebec G5L 3A1 Canada}
  \email[Correspondence email address: ]{ludovic.moreau@univ-grenoble-alpes.fr}

%\corau{*Corresponding author: ludovic.moreau@univ-grenoble-alpes.fr}

\begin{abstract}
Sea ice mechanical properties are involved in dynamical processes acting from the scale of meters to several hundred kilometers. The current rapid changes in the state of polar sea ice require a better understanding and modeling of these processes and, therefore, accurate measurements of properties including sea ice thickness, density, Young's modulus and Poisson's ratio. 
These properties can be measured by tracking the propagation of elastic waves within the ice. Recent technological advances have enabled the use of fiber-optic cables as cost-effective, dense seismic arrays. Once connected to an interrogator unit and mechanically coupled to a medium, here the ice cover, these cables can monitor strain field propagation, using a technique called Distributed Acoustic Sensing (DAS). In this work, we describe the use of such an array of sensors in the coastal ice of the St.~Lawrence Estuary, Canada, where a 600~m long optical fiber was deployed across three different morphological sea ice conditions. 
During hour-long recordings, we measured the propagation of both multi-modal seismic signals generated by active sources and hydro-elastic swell. We computed dispersion curves of active signals and used Continuous Wavelet Transform (CWT) to observe the evolution of swell characteristics in the different ice areas. The dispersion curves were successfully inverted to measure the spatial evolution of ice thickness, and Young's and flexural rigidity in each of these areas. We observed ice thicknesses from 25~cm to 68~cm and Young's modulus values between 4.5~GPa and 5.7~GPa, in good agreement with values derived from collocated geophone arrays and drill hole thickness measurements. DAS systems therefore appear to be effective in evaluating heterogeneous sea ice mechanical properties and thus sea ice formation history and dynamics.

%By following the propagation of both swell and active noise along the fiber, we succeeded in retrieving
\end{abstract}

\maketitle

\section{Introduction}

The thin ice cover that floats on the surface of the polar oceans is continuously subjected to stresses imposed by the winds and the underlying ocean. Its mechanical properties (density, thickness, elastic (or Young's) modulus, and Poisson's ratio) control how it deforms in response to these forcings across a wide range of spatial scales.

At large scales, from tens to hundreds of kilometers, ice deformation is of central importance for regional and global climate predictions. Over the past half-century, a variety of continuum rheological models have therefore been developed to represent sea ice deformation. Some describe sea ice as a viscous-plastic medium \citep[e.g.][and references therein]{hibler1979}, others as an elastic-decohesive material \citep{Schreyer2006}, and still others as a visco-elasto-brittle medium \citep{dansereau_maxwell_2016}. Despite their differences, all these approaches rely on knowledge of the value and spatial distribution of some, if not all, of the aforementioned mechanical properties.

At smaller scales, ranging from a few meters to a few kilometers, deformation is dominated by processes such as fracturing and ridging. In particular, at these scales, surface gravity waves generated by winds in the open ocean can induce hydroelastic deformation of the ice cover. Depending on both the ice mechanical properties and the characteristics of the waves \citep{mokus_swiift_2026, thomson_wave_2022}, this forcing may lead to a widespread breakup of the ice into a collection of fragments of varying sizes and thereby gives rise to a highly dynamic region known as the Marginal Ice Zone \citep[MIZ, ][]{squire_ocean_2020, dumont_marginal_2022}. Recent laboratory works \cite{saddier2024breaking, Auvity2026} emphasized the role of physical parameters and material properties on the wave-induced breakup of mimetical materials.

Sea-ice dynamics across all these scales therefore depend critically on its mechanical properties. Like any continuous floating ice sheet, sea ice supports the propagation of three types of guided mechanical waves: shear, longitudinal, and flexural. The speeds and dispersive characteristics of these waves are governed by the density, thickness, Young's modulus, and Poisson's ratio. As a result, these effective mechanical properties can be inferred from measurements of mechanical wave propagation within the ice cover.
\smallskip

Active seismology provides a non-invasive method for extracting key information for a range of fundamental and practical questions related to sea ice. Early experiments, such as the ones of \cite{hunkins_seismic_1960} and \cite{stein_inversion_1998}, required human presence on the ice and so limited the ice conditions that could be sampled and the frequency of sampling.
%Such active seismic measurements rely on the artificial generation of elastic waves which requires a human presence on the ice. The harsh conditions of the Arctic environment make these kind of measurements on arctic sea ice quite rare. 
%
In the early 2000's, new methods were developed to measure the physical characteristics of a medium from the recording of ambient seismic noise. \cite{lobkis_emergence_2001} showed that the Green's function between two acoustic sensors can be derived from the computation of a long time-average cross-correlations. Using this method, \cite{shapiro_emergence_2004} successfully retrieved dispersion curves of Rayleigh waves at the Earth's surface. 

Later, advances in seismic sensor battery capacity enabled the application of this passive methodology under the harsh conditions typical of polar environments, including multi-day recordings of sea ice seismic noise.
Using cross correlation methods between three different seismic stations separated by more than 30 kilometers, \cite{marsan_sea-ice_2012,marsan_characterizing_2019} measured dispersion curves of large scale seismic wavefields. 
\begin{comment}
Two main vibration modes were detected, corresponding to the non-dispersive horizontal shear mode and the dispersive flexural mode. These two modes appeared unsynchronized, implying that the related wavefields had been generated by different processes. Ocean swell was identified as the main source of the observed flexural mode. Yet, the low-frequency dispersion curve of the flexural mode does not allow the inversion of a precise effective ice thickness at the kilometer scale.
\end{comment}
% \cite{marsan_sea-ice_2012} succeeded in extracting sea ice thickness thanks to the measurement of swell propagation in ice and correlation between seismic sensors. \cite{sutherland_observations_2016} also managed to measure swell and wind-generated waves dispersion relation in ice as well as their attenuation. \seb{using accelerometers IMU, describe quickly.}
%
\cite{moreau_sea_2020} later deployed a much denser array of 247 geophones spanning approximately 250 meters on first-year sea ice in Svalbard. Using specific array processing methods and noise correlation functions, dispersion relations for longitudinal, shear, and flexural waves were extracted from ambient seismic noise \citep{serripierri_recovering_2022}, allowing the inference of ice elastic properties and thickness.
% While a large number of sensors were used for to get ice characteristics from the waves dispersion, \cite{moreau_accurate_2020} also proposed an estimate using a reduced number of sensors. Using 3 to 5 sensors, the ice plate properties could be directly inverted from the icequakes waveform.
\cite{moreau_accurate_2020, moreau_analysis_2023} also used machine learning methods to invert the waveforms of tidal icequakes for monitoring of the spatial and temporal evolution of fast ice properties with a much reduced number of receivers (3 to 5).

\smallskip
These so-called passive methods ease the deployments of seismic sensors arrays, as fieldwork is required only for deployment and recovery stages,  and enable retrieving sea ice mechanical properties directly from long-time recording of natural seismic noise. Yet, the precise inversion of these properties still requires either a dense array of seismometers or the hypothesis of material homogeneity between two sensors. Characterizing \textit{spatial variations} of mechanical properties still requires a dense array of seismometers, deployed over several hundred meters, making such studies quite expensive and logistically burdensome.

%the deployment of several ponctual sensors, which should be \seb{Limited by Nyquist spatial frequency, therefore deployment of such arrays over several hundred meters in order to get sea ice spatial variability need heavy logistics.}

% Development of Distributed Acoustic Sensing. How does it work ? Since when ? For which purpose ? Progressive interest of DAS to measure ice mechanical properties than sea ice properties.

Recent advances in telecommunication and in the ubiquity of fiber-optics cable prompted the emergence of new measurements using Distributed Acoustic Sensing (DAS) systems. DAS measurements use a fiber-optic cable coupled to an optoelectronic interrogator unit in order to measure the strain field acting on the cable. These methods take advantage of Rayleigh scattering caused by defects and refractive index inhomogeneities in the fiber; the optoelectronic interrogator sends laser light pulses through the optical fiber that are partly backscattered along the whole fiber length. Using optical interferometry, the DAS system measures phase changes of the backscattered signal, which are related to strain distortion of the fiber. Subsequently, a time-to-distance conversion enables the measurement of the strain rate signals at different spatial regions of the fiber, transforming the cable into a dense array of sensors. The strain rate signal is computed over fiber sections for which the length, called the gauge length, is typically on the order of meters \citep{lindsey_broadband_2020}. Given the mechanical coupling between the cable and its surrounding material, DAS can constitute a dense linear array sensing material deformation over several kilometers.

DAS measurements system have been deployed for various purposes, including earthquake detection, ambient noise interferometry in urban areas, and structural health monitoring \citep{lindsey_fiber-optic_2021}. DAS has been increasingly used in cryospheric applications including permafrost imaging and glaciers monitoring \citep{wagner_permafrost_2018, booth_distributed_2020}. Unused telecommunication fibers (also known as dark fibers), deployed onshore or on the seafloor can also be used to perform acoustic sensing. \cite{baker_rapid_2022} observed the refreezing of a MIZ using seafloor cable in the Beaufort Sea by comparing swell dispersion curves between an ice-free and ice-covered ocean state. Using the same seafloor cable, \cite{smith_observations_2023} monitored MIZ ice edge motion over several days and quantified surface swell attenuation along the cable.

%\ludo{somehow we need to cite Marsan et al (2019 -> Characterizing horizontally-polarized shear and infragravity vibrational modes in the Arctic sea ice cover using correlation methods) We also need to cite Nziengui et al (2022} -> Measuring Floating Ice Thickness with Distributed Acoustic Sensing: a Case Study on a Frozen Mountain Lake.

The potential application of DAS systems for ice surface seismicity have been explored through several studies \citep{nziengui-ba_measuring_2023, xie_ice_2024, quinn_freshwater_2025}. \cite{nziengui-ba_measuring_2023} were able to measure the ice thickness and Young's modulus of lake ice from the monitoring of both ambient seismic noise and signals generated by active sources. \cite{xie_ice_2024} and \cite{quinn_freshwater_2025} also succeeded in inferring Young's modulus or ice thickness of frozen lakes from the propagation of artificially generated flexural waves along a fiber-optic cable.

These latest studies show that DAS system can be efficiently used to estimate and monitor ice elastic properties, conditional on adequate mechanical coupling between the cable and ice. Yet, to our knowledge, spatial variations of these properties along the deployed cable have not been studied in depth. In this study we propose to take a step further and verify how a DAS system can measure spatial variations of sea ice mechanical properties in an inhomogeneous sea ice field. For this, we perform seismic measurements over a 600-m long distance of a continuous yet heterogeneous sea ice plate. By coupling the fiber-optic cable to the ice with plastic boxes and a thin layer of snow, we obtain a cost effective way to measure the propagation of sea ice strain rate fields. Despite the high energy consumption required to perform DAS, we are able to capture both hydro-elastic swell propagation and seismic signals generated by active sources during hour-long recordings on two consecutive days. Using both of these signals, we are able to measure with good precision the spatial evolution of sea ice mechanical properties across the different ice regions. Our results are in agreement with observations from independent seismic recording using geophone arrays as well as with drill hole thickness measurements, providing confidence that DAS systems can be used to efficiently record and understand the history of sea ice.

% Due to absence of section numbers, we have re-written the last paragraph.
The paper is organized as follows. We first present the study site and sea ice conditions and describe the experimental set-up, instruments and protocols for seismic noise generation and data acquisition. Then, we present the signal analysis methods and results obtained from the detection of passive waves and active sources. Finally, we discuss how DAS estimates compare with independent observations and how they relate to the sea ice cover characteristics.

% Old Version of the last paragraph
%The paper is organized as follows. Section~\ref{sec:methods} presents the study site and sea ice conditions and describes the experimental set-up, instruments and protocols for seismic noise generation and data acquisition. Section~\ref{sec:result} presents the signal analysis methods and results obtained from the detection of passive waves and active sources. Finally, section~\ref{sec:discussion} discusses how DAS estimates compare with independent observations and how they relate to the sea ice cover characteristics.

\section{Study site and methods}
\label{sec:methods}

%\begin{figure}[ht!]
%    \includegraphics[width=1.0\textwidth,,trim={1.5cm 3.5cm 1.5cm 3.5cm},clip]{Figures/fig_map_bdhh_das2026.png}
%    \caption{Bathymetry of Baie du Ha!Ha! relative to the tidal water chart datum, as obtained from the 10-m resolution non-navigational data product of the Canadian Hydrographic Service (CHS NONNA10). The black solid line shows the location of the optical fiber and the dash line the ice area shown in \dy{b}.}
%\label{fig:haha_bathymetry}
%\end{figure}
\subsection{The BicWin campaign in Baie~du~Ha!~Ha!, St.~Lawrence Estuary}

\begin{figure*}[ht!]
    \includegraphics[width = .95 \textwidth,,trim={0cm 0cm 0cm 3.5cm},clip]{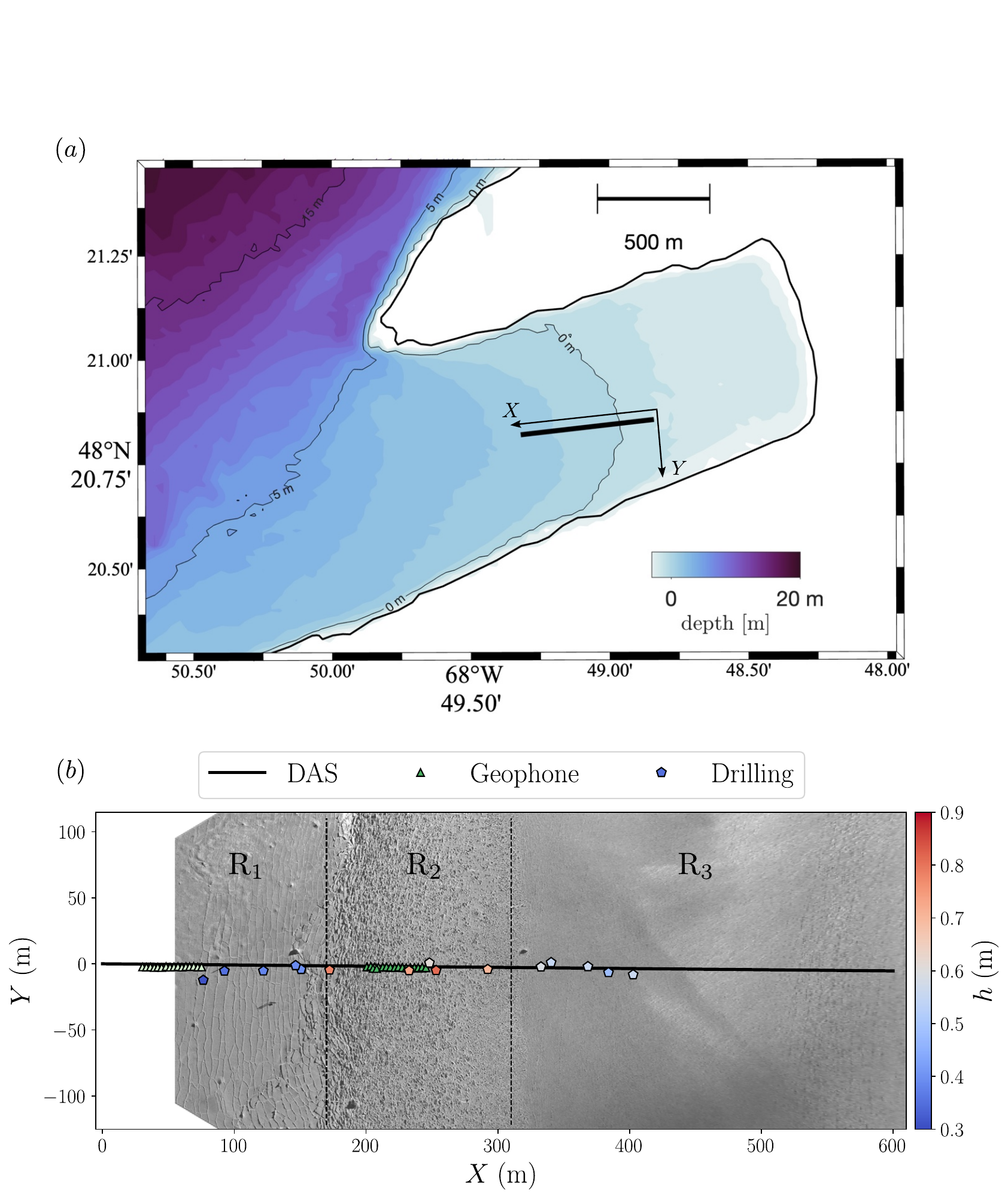}
    \caption{ (a) Bathymetry of Baie du Ha!~Ha! relative to the tidal water chart datum, as obtained from the 10-m resolution non-navigational data product of the Canadian Hydrographic Service (CHS NONNA10). The black solid line shows the location of the optical fiber, coordinate system $(X,Y)$ used in (b) is also represented and accurately oriented. (b) A georectified aerial picture of Baie du Ha!~Ha! taken on 5 February 2025 at an altitude of 111~m overlaid with drill hole thickness and geophones measurements  made on 10 and 11 February, the day the 600-m long optical fiber was installed (black line). The cartesian coordinate system is associated with the DAS fiber with the origin located at one end and the fiber at $Y=0$. Vertical dotted lines separate three regions with different morphologies and histories.}
    \label{fig:situation_map_bathymetry}
\end{figure*}

This study was conducted in Baie~du~Ha!~Ha!, in the Bic National Park located on the south shore of the St.~Lawrence Estuary near Rimouski, QuÃ©bec, Canada. The rectangular bay is 2~km long and 1~km wide and opens approximately to the west (Figure \ref{fig:situation_map_bathymetry}a). It is located downstream of a sensible heat polynya that is favorable for wind-wave generation during winter. From January to March, cold air temperatures and westerly winds cause waves to grow and sea ice to form, drift and accumulate along the shore and into the bay. As waves are aligned with the bay's longitudinal axis, this makes it an ideal site to study wave-ice interactions, which has been done annually since 2016 \citep[e.g.,][]{sutherland_marginal_2018, Galbraith2024}. %\seb{add ref old Bicwin or freezing of St.~Lawrence}. 

Data presented here were obtained during the 2025 BicWin campaign in February 2025, using methods that were developed and tested during the 2024 BicWin campaign \citep{kuchly_integrated_2025}. We used unmanned aerial vehicles (UAV), accelerometers, and two types of seismic sensors to simultaneously record the propagation of hydro-elastic waves forced by incoming ocean surface gravity waves ($T<5$~s) and actively generated mechanical waves to estimate sea ice mechanical properties. The seismic sensors include a 600-m long single-mode fiber-optic cable coupled to a DAS interrogator unit, described in section~\ref{sec:methods_das}, and an array of geophones (see section~\ref{sec:methods_geophones}). 

\subsection{Sea ice conditions}
\label{sec:method_ice}

%\seb{Describe sea ice condition, three different areas (use situation picture to help), describe ice ridges due to ice fracture and ice formation (frasil compaction). Define different ice regions $R1$, $R2$, $R3$}

The orientation of Baie du Ha!~Ha! favors a diverse and rapidly changing icescape thanks to frequent wave-ice interaction events occurring under varying weather and tidal conditions. Figure~\ref{fig:situation_map_bathymetry} (b) shows sea ice conditions on 5 February 2025 at the place where the optical fiber was deployed on 10 February. From the shore side of the bay (to the right) up to the ice edge, three distinct ice regions can be identified, each formed at different times under different conditions. Region 1 (called R$_1$, $X<$170~m) is the one that formed the first. It is a relatively smooth sea ice region with an average thickness $h=$36~cm likely formed by the compaction of frazil or slush against the ice edge by waves, which eventually consolidated during subsequent calm and cold conditions. The darker lines predominantly perpendicular to the bay longitudinal axis are fractures likely caused by waves on January 21 which refroze later. 
%\seb{From meteorological data, those fractures might have been formed during intense westerly wind events, on January 27.} \dy{[It would be nice to say approximately when did these event happened]}. 
Due to the shallowness of the bay at this location (Fig. \ref{fig:situation_map_bathymetry}a), waves can only interact with sea ice around high tides, since most of their energy is dissipated before reaching this point at low tide. The bumps one can see in this region and elsewhere are caused by boulders sitting on the sea floor that push, fracture and deform the ice sheet from below at low tide when the water depth is nearly zero. In region 2 (R$_2$, 170~$<X<$~310~m), the ice is a rough jumble of ice fragments with irregularities about 20~cm high which may have resulted from the fragmentation and radiative push from much larger waves causing rafting. The level of deformation is higher near the former ice edge and decays outward, suggesting progressively increasing compression induced by the convergent wave radiative stress applied in this region ice as waves attenuated. The $e$-folding damping scale of a few hundred meters and the average thickness of the ice jumble ($h=$72~cm) is consistent with observations made in this area \citep{sutherland_marginal_2018}. At the time of the picture in Fig. \ref{fig:situation_map_bathymetry}b, the jumble was completely refrozen, with a thin snow layer hiding the smaller irregularities.
The third region (R$_3$, $X>$~310~m) had a smooth surface and an average thickness $h=$54~cm, in contrast to the other two regions. Ice there was likely formed around 27 January during cold weather and moderate winds. %\steph{Check age of ice in region $R_3$ with the Pic Champlain Camera pictures}. \seb{Pic Champlain Camera records start only on the 29 January... My guess is that the ice region R$_3$ was formed on January 21, based on meteorological data (-20$^{\circ}$C in Rimouski)}. 

%Several local measurements of sea ice thickness have been performed along the optical fiber (see Figure~\ref{fig:situation_map}). At each location, a hole is drilled into the ice, and we measure the thickness with a stick hooked at the base of the ice. A total of 15 manual thickness measurements have been performed along the fiber. From these measurements, we observe a spatial variation of the ice thickness between the different ice conditions R$_1$, R$_2$ and R$_3$. From the bottom of the bay to the ice edge, the first ice area R$_1$ is about $36$~cm thick. The ice jumbled region R$_2$ is on average 72~cm thick, twice the thickness of the region R$_1$, which suggests that this specific area might have thickened through break-up and compressive rafting caused by storm waves and the associated (compressive) radiative stress. The mean thickness of the last region R$_3$ is $54$ cm, which is thicker than R$_1$, suggesting that ice found in R$_1$ and R$_3$ have been formed during different cold events.

\subsection{DAS system deployment}
\label{sec:methods_das}

%\seb{Orientation of fiber with respect to ice, interrogator unit connected to mobile electric power supply (acquisition of about 1h10). Deployed 3 different days at high tide. Fiber clamped to ice every 20 meters thanks to cellphone cases. Thinn layer of snow during night of 10th to 11th of February.}

The 600 meter long optical fiber was deployed on 10 February 2025 on the ice in the middle of Baie du Ha!~Ha! recovered two days later on 12 February. The tight-buffered flat drop cable equipped with 2 single-mode fibers was deployed orthogonal to the main ice fractures and ridges characteristic of the prior history of ice and its interaction with incoming waves. Every 20 meters, we clamped the cable against the ice surface by screwing a plastic plate on top of the fiber to the ice. The GPS position of the optical fiber depicted in Figure~\ref{fig:situation_map_bathymetry}b was recorded every 20 meters, at these specific positions. After a snowfall event overnight on 10 February, the optical fiber was covered by a thin layer of snow, about 3~cm thick on average, which improved the mechanical coupling with ice. Despite our best efforts, some parts of the fiber remained loose and uncoupled to the ice as can be seen from the high-energy bands in the spatio-temporal plot displayed in Figure~\ref{fig:DAS_spatio}a. Denoting $x$ the curvilinear coordinate along the fiber, these uncoupled intervals are $170$~m~$<x<240$~m and $270$~m~$<x<310$~m, which coincide with the crossing of the most irregular ice region $R_2$.

We successively performed DAS observations on the 10, 11 and 12 February. We connected the optical fiber to a Febus A1 interrogator unit powered by a mobile power supply, enabling continuous acquisition for one hour. We placed the interrogator unit and its power supply in a plastic case allowing protection against wind and snow. The interrogator and the mobile power were recovered daily. We hereafter only present observations of the 11 and 12 February, for which coupling with ice was improved by freezing and snow.

The fiber-optical cable was sampled with a time acquisition frequency of $84$ Hz on the 11 February and of $840$ Hz on the 12 February. The gauge length, equivalent here to the inverse of the spatial sampling rate, was set to $2.0$ meters on 11 February and to $1.0$ meter on 12 February.

%\begin{figure}[ht!]
%    \includegraphics[width = 1. \textwidth]{Figures/situation_picture_with_ice_labels.pdf}
%    \caption{A georectified aerial picture of Baie du Ha!ha! taken on 5 February 2025 at an altitude of 111~m overlaid with drill hole thickness measurements made on 10 and 11 February, the day the 600-m long optical fiber was installed (black line). The cartesian coordinate system is associated with the DAS fiber with the origin located at one end and the fiber at $Y=0$. Vertical dotted lines separate three regions with different morphologies and histories.}
%    \label{fig:situation_map}
%\end{figure}

%\seb{Interesting pictures and orthophoto 0210/Drones/bernache/06-ortho_003 ; 0210/Drones/bernache/09-ortho_004 ; 0210/Drones/mesange/01-doc_001 ; 0210/Drones/mesange/07-doc_004}

\subsection{Geophone deployment and thickness measurements} \label{sec:methods_geophones}
%\seb{Description of geophones used, refer to BicWin24. Precise which deployments we are using (select date and place in HaHa! Bay). }

%Several ponctual measurements of the sea ice thickness have also been performed along the optical fiber. (See figure \ref{fig:situation_map}). At each location, a hole is drilled into the ice, enabling to submerge a measuring tape. Once we hook the bottom edge of the hole, we denote the ice thickness and GPS location. A total of $15$ manual thickness measurements have been performed along the fiber.  

We deployed two linear arrays of 16 geophones along the single-mode optical fiber. The first array was deployed on a smooth ice cover (in R$_1$) and the second array in a rougher ice area (in R$_2$, see Figure~\ref{fig:situation_map_bathymetry}b). For both deployments geophones were placed every 3~m along the fiber such that the distance between the first and last geophone was 45~m. We followed the protocol developed by \cite{kuchly_integrated_2025} to measure the sea ice effective Young's modulus, Poisson coefficient, ice thickness and density from the multimodal propagation of elastic waves. Briefly, we excited mechanical waves of three types by hitting on the ice surface 5, 8 and 11 meters away from the first and last geophones of the array. How we actively generate the desired seismic noise is described in section~\ref{sec:methods_sources}. Overall, this method enables a quick, reproducible and efficient measurement of all the relevant effective mechanical properties. Finally, the ice thickness was measured at multiple locations along all three regions using a meter stick or tape through drill-holes (see Fig.~\ref{fig:situation_map_bathymetry}b).
%At each of these 6 locations, we generate the three types of waves that can propagate in a thin solid plate : shear waves, longitudinal waves and out-of-plane waves, denoted respectively E (for east), N (for north) and Z for vertical). The north direction corresponds to the direction of the linear array. These three types of waves are generated by following the "ZEN" procedure. First, out-of-plane waves (Z) are generated by a single person jumping three successive time on the ice. Each jump is delayed from the previous one by a few seconds to allow the mechanical waves to propagate along the array. Then, shear waves (E) are generated by hitting the ice with a hammer, in a direction orthogonal to the array. Three successive hits are performed, separated by a few seconds. Finally, longitudinal waves (N) are created by three consecutive hits in the ice with a hammer in the direction of the array. 
%The results obtained from the three deployments are described in section~\ref{sec:result}. 

%\seb{What should we do with the tomography ?}
%\steph{Ludo ?}

%\subsubsection{HEADING}

\subsection{Active and passive sources}
\label{sec:methods_sources}
%\seb{Description of how we perform active sources. (ZEN method every 20 meters). Description of waves condition : regular swell observed on the 11th of February. Using wind data ? Swell was observed on the 12th, but its intensity is rather low on the 10th, associated to low coupling of fiber wth ice.}

\begin{figure*}[ht!]
    \includegraphics[width=1.0\textwidth]{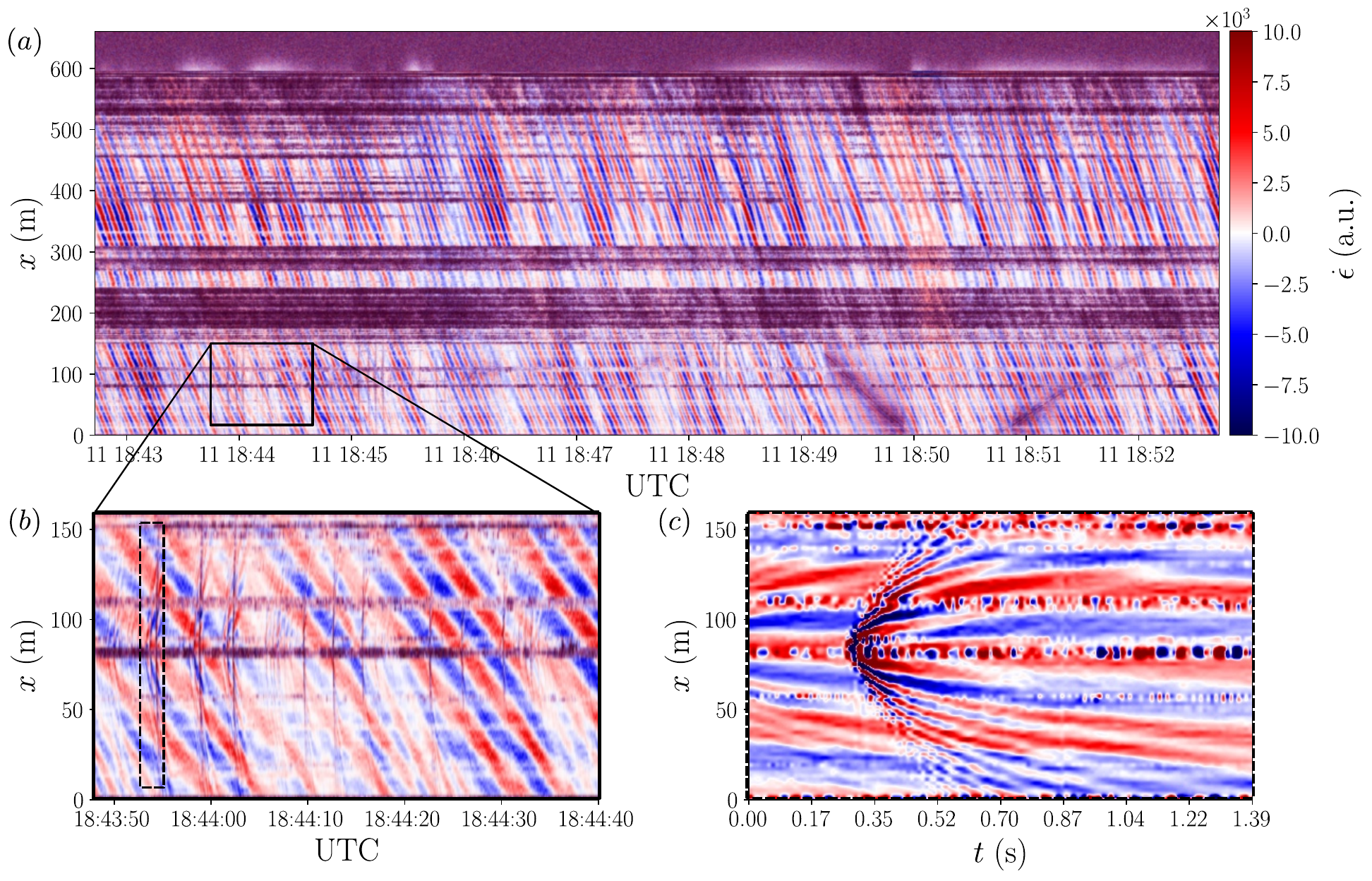}
    \caption{a) A 10 minute-long space-time plot of the strain rate signal measured on 11 February 2025. b) Zoom on the space-time strain rate signal corresponding to active noise generation following the ZEN procedure. c) Zoom on the space-time strain rate signal related to the generation of a flexural out-of-plane wave (Z) at $X=$83~m.}
    \label{fig:DAS_spatio}
\end{figure*}

During our measurements, we recorded both the ambient seismic noise and seismic signals generated by active sources.
Here, the main passive strain rate signal we recorded during the two-day experiment (11 and 12 February) was generated by incoming ocean surface gravity waves (swell) exciting hydroelastic out-of-plane waves in the ice sheet \citep{stein_inversion_1998} that propagated from the ice edge shoreward (Fig.~\ref{fig:DAS_spatio}a). These waves have frequencies in the range 0.1 to 0.7~Hz and were characterized by a narrow spectrum with a peak frequency $f_p$ of 0.21~Hz on the first day (11 February) and 0.29~Hz on the second day (12 February).
%The propagation of these hydroelastic waves depend on the sea ice mechanical properties, and can be used to invert these properties as presented in the Results section. 

In addition to naturally occurring hydro-elastic waves, we also actively generated other types of mechanical waves with frequencies from 1 to 40~Hz. The method used to inverse ice mechanical properties from the multimodal propagation of mechanical waves through a linear array of geophones can be applied to DAS. By hitting the ice every 20 meters along the cable, we could also measure the propagation of waves along the optical fiber axis and thus invert sea ice mechanical properties over the entire cable length.

At each active source location, we generated the three types of waves that can propagate in a thin solid plate: shear waves, longitudinal waves and out-of-plane waves, denoted respectively E (for east), N (for north) and Z (for vertical). The N direction corresponds to the axis orientation of the linear array, Z to the vertical, and E the direction orthogonal to the array. These three types of waves are generated following the "ZEN" procedure. To generate out-of-plane (Z) waves, a person jumps on the ice three times, to create replicates. A few seconds delay between each jump allows the mechanical waves to propagate over the entire array length before the next waves are generated. Shear waves (E) are generated by hitting the ice with a hammer, in a direction orthogonal to the array. Again, three successive hits are performed. Longitudinal waves (N) are also generated using the hammer, but hitting instead in the direction of the array. 

Geophones can detect and measure the three modes, as they measure the three-dimensional velocity of the ice. The DAS system is only sensitive to the strain along the fiber axis such that only the longitudinal mode (N) and the out-of-plane mode (Z) induce an elongation of the optical cable. These two modes are sufficient to invert for two mechanical properties along the optical fiber. Assuming the Poisson coefficient to be constant, we choose to invert both ice effective Young's modulus and effective thickness on the 600 meters deployment. 

Strain rate signals associated with the active noise can be observed in Figure \ref{fig:DAS_spatio}b), with the propagation of the out-of-plane mode observed in detail in Figure \ref{fig:DAS_spatio}c). Between each source point, a set of sensors can be defined to track the propagation of the different modes and invert for the local properties of the ice. 

We also used these active sources to correctly map the curvilinear coordinate given by the interrogator unit to real world, spatial coordinates. We compared the x-position of observed signals on spatio-temporal plots (Figure \ref{fig:DAS_spatio}) with the distance of the corresponding source point to the interrogator unit position. By doing so, we observed that the relative error between curvilinear coordinate given by the interrogator and the real spatial coordinates is of the order of 1\%. We therefore decided to neglect this correction.

\section{Results}
\label{sec:result}

\subsection{Young's modulus and thickness inversion from active sources}
Figure \ref{fig:DAS_spatio}c shows a space-time strain rate record from the DAS interrogator after the generation of Z waves from a person jumping. Waves propagation was typically observed over 50-100 m, enabling application of the process described by \cite{kuchly_integrated_2025} to extract flexural and longitudinal dispersion curves. This process combines the wave fields recorded from several sources aligned with the same linear array to compute the frequency-wavenumber spectrum. However, unlike in previous work, we can use a single source and measure propagation along three fiber segments of length $L$ subsequently separated by 3~m. As $L = 40$~m, the dispersion curves we obtain for each source correspond to waves recorded from 3 to 49~m from the source point. The longitudinal wave dispersion curve was inverted to estimate the Young's modulus of the ice using \citep{stein_inversion_1998}
\begin{equation}
E = \rho(1-\nu^2)\frac{k_L^2}{\omega^2},
\label{eq:EfromkL}
\end{equation}
where $\rho$ is the density of the ice. The estimated Young's modulus was then used to infer ice thickness by fitting the flexural wave dispersion curve with that given by Eq.~\ref{eq:shallow_hydroelastic_waves} below. This procedure was applied to all sources along the optical fiber, enabling a spatial resolution of about 20~m for both Young's modulus and ice thickness, as shown in Figure~\ref{fig:EhD_vs_x}.

\subsection{Hydroelastic waves excited by swell}
\label{sec:hydroelastic}

Hydroelastic waves generated by surface gravity waves (or swell) propagate from the ice edge shoreward and induce a continuous, cyclic deformation of the sea ice plate. The DAS system is highly sensitive to these deformations, and the propagation of these mechanical waves from one fiber end ($x=$600~m) to the other is quite clearly observed (Fig.~\ref{fig:DAS_spatio}a). The propagation of hydro-elastic waves within a floating plate is dispersive, and follows a relation that depends on physical properties of both water and sea ice \citep{stein_inversion_1998, domino_dispersion-free_2018,Ledoudic2025}, given by
%The wavenumber $k$ is related to the wave angular frequency $\omega$ through the following relation :
%\seb{How can I explain that hydroelastic waves are similar to out-of-plane QS mode ? But I use a simplified dispersion relation.}
%
\begin{equation}
    \omega^2 = \left(gk + \frac{D}{\rho_w}k^5 \right)\tanh(Hk)
    \label{eq:shallow_hydroelastic_waves}
\end{equation}
where $k$ is the wavenumber, $\omega$ is the angular frequency, $g$ is the gravitational acceleration, $\rho_w$ the seawater density and $H$ the water depth. The coefficient $D=\frac{Eh^3}{12(1-\nu^2)}$ is the ice flexural rigidity, with $E$ the ice Young's  modulus, $h$ the ice plate thickness and $\nu$ the Poisson's ratio.

\begin{figure*}[t!]
    \includegraphics[width = 1. \textwidth]{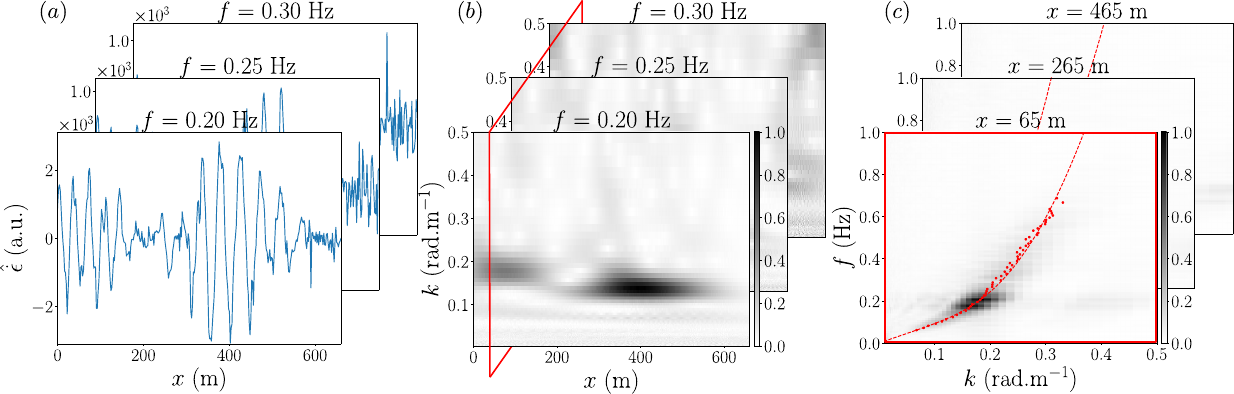}  
    \caption{Illustration of the inversion process for obtaining the sea ice flexural rigidity $D$ from the hydro-elastic signal induced by swell. First, the strain rate signal is split into each frequency component (a). Then a Continuous Wavelet Transform (CWT) is computed for each frequency component, leading to a scaleogram for each frequency (b). Finally, a section of all scaleograms (shown in red in panel b) for a given position $x$ is selected in order to get the space-time spectrum at this position (c). The hydro-elastic dispersion relation is then obtained by fitting the maxima of the space-time spectrum (red dots) for each position along the fiber.}
    \label{fig:CWT_explanation}
\end{figure*}

Variations of the flexural rigidity along the fiber imply a continuous evolution of the dispersion relation that can be determined using the Continuous Wavelet Transform (CWT) method. First, we filter out the signal $\dot{\epsilon}(x,\omega)$ related to a single angular frequency component $\omega$ of the wave spectrum. This filtered signal presents various wavelength values along the fiber. Figure~\ref{fig:CWT_explanation}a shows an example of the filtered signal $\dot{\epsilon}(x,\omega)$ at the frequency $f = 0.20$~Hz. We observe wavelengths ranging from 35~m in R$_1$ to 45~m in R$_3$.

We then apply the CWT to localize a signal simultaneously in the space $x$ and wavenumber $k$ domains. This enables tracking the evolution of the hydroelastic wavenumber along the curvilinear coordinate $x$ for a given frequency $f$. The CWT involves computing the convolution product of the filtered signal $\dot{\epsilon}(x,\omega)$ with a complex valued window function $\psi_{s,x_0}(x)$, called a wavelet. This window function is a translated and rescaled version of a single function $\psi(x)$ called the mother wavelet, such that $\psi_{s,x_0}(x) = \frac{1}{\sqrt{s}}\psi \left(\frac{x - x_0}{s}\right)$, where $x_0$ is the translation coefficient and $s$ a scaling factor. Therefore, the convolution product, given by
\begin{equation}
    \hat{\dot{\epsilon}}(s,x,\omega) = \frac{1}{\sqrt{s}} \int_{\infty}^{\infty} \dot{\epsilon} \left( u,\omega \right) \psi^* \left( \frac{u-x}{s} \right) \, du,
\end{equation}
enables the creation of an image, called a scaleogram, with components that show how well the signal matches a given scaled version of the mother wavelet $\psi$ and the evolution of this match with the coordinate $x$. 
In order to increase the compatibility between filtered signal and wavelets, we choose a complex Morlet mother wavelet $\psi(x) = \frac{1}{\sqrt{\pi B}}\mathrm{e}^{-\frac{x^2}{B}}\mathrm{e}^{i2\pi Cx}$ with $B = C = 1.5$. This choice represents a trade-off between spatial and spectral width that optimally captures the wavelength variations in the strain rate signal.

The scaleogram computed from the filtered signal with $f=$0.20~Hz is shown in Figure~\ref{fig:CWT_explanation}b. We use the Python package \texttt{PyWavelet} to perform the CWT and to convert each scale $s$ to the corresponding wavenumber $k$, the so-called scaleogram. The wave propagation direction relative to the optical fiber axis is taken into account in this correspondence between scales and wavenumbers. Using a beamforming method \citep{rost_array_2002}, we found a relative angle between swell direction and the fiber axis of 13.5$^\circ$ and 8.5$^\circ$ on the 11 and 12 February, respectively. The real wavenumber $k$ is then computed from this angle and the wavenumber $k^*$ observed from the strain rate signal, using geometric projection $k = k^*/\cos{(\theta)}$. 
The scaleogram shown in Figure~\ref{fig:CWT_explanation}b presents two different wavenumber values, i.e. $k=$~0.18~rad~m$^{-1}$ and $k=$~0.14~rad~m$^{-1}$ each associated to ice regions R$_1$ and R$_3$.
%ranging between the fiber beginning tip to $x = 170 \; \mathrm{m}$ and between $x = 300 \; \mathrm{m}$ and $x = 500 \; \mathrm{m}$. These fiber sections correspond to the ice regions  we identified previously. 

Scaleograms computed for each angular frequency component of the wave frequency spectrum are obtained at each position of the fiber. As each scale $s$ corresponds to a wavenumber $k$, we build a $k$-space-time spectrum $|\hat{\dot{\epsilon}}|(k,x,f)$ that characterizes the relative importance of each wavenumber and frequency components $(k,f)$ in the signal recorded at $x$. We compute several space-time spectra over minute-long signal windows, and average these spectra to get the final space-time spectrum $|\hat{\dot{\epsilon}}|(k,x,f)$. This spectrum is shown in Figure~\ref{fig:CWT_explanation}c for $x=$~65~m.

For each position $x$ of the space-time spectrum, we extract the maxima location, represented by red dots in Fig.~\ref{fig:CWT_explanation}c, using an order 2 polynomial fit. Then, the positions of these maxima in the plane ($k$,$f$) are fitted by the shallow water dispersion relation of hydroelastic waves (Eq.~\ref{eq:shallow_hydroelastic_waves}). Using bathymetric and local tidal elevation data, we compute the water depth below the optical fiber at the instant of the measurement, in order to precisely infer the ice flexural rigidity $D$ along the optical fiber. 

% Three space-time spectra and associated dispersion relation are represented in figure \ref{fig:flexural_fit}. Each spectrum corresponds to a different sea ice condition depicted previously. For $x = 100 \; \mathrm{m}$, located in the ice region $R_2$, made of jumbled ice, we observe a steeper dispersion relation, correlated with a higher value of the flexural rigidity $D$, which is twice the value of measured modulus at $x = 100 \; \mathrm{m}$, in the smooth ice region $R_1$.  

\begin{figure*}[t!]
    \includegraphics[width=1.\textwidth]{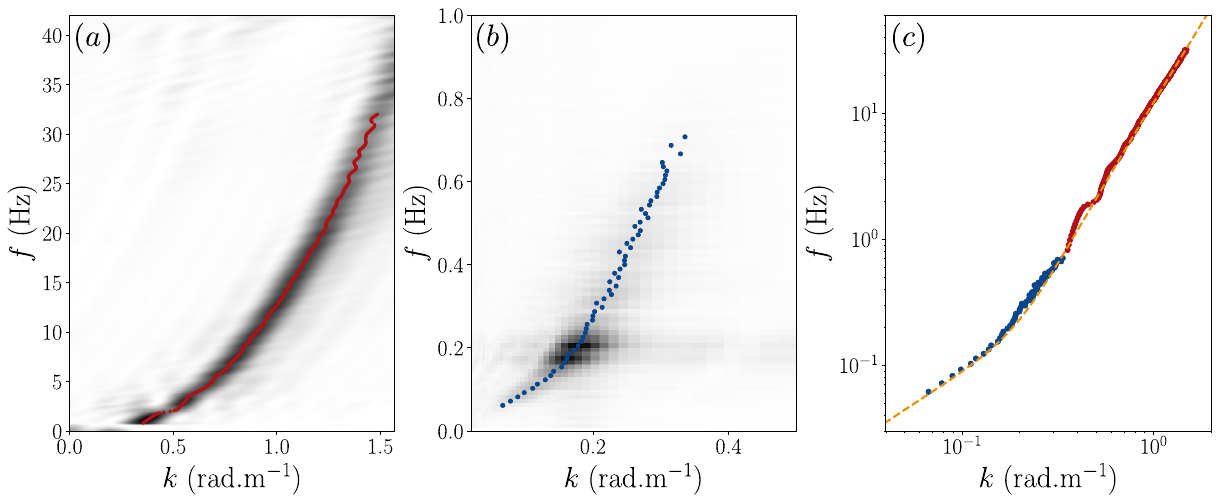}
    \caption{Space-time spectra $|\hat{\dot{\epsilon}}|(k,f)$ obtained at $x=$~80~m from active seismic (a), and passive hydroelastic signals (b). Subpixel position of each spectrum maxima are detected and represented as color dots. Combination of these two sets of $(k,f)$ components are shown in (c). The dash line represents the hydro-elastic dispersion relation fitted in the $(k,f)$ space obtained from active seismic noise (blue dots).}
    \label{fig:FK_active_passive}
\end{figure*}

In Figure~\ref{fig:FK_active_passive}, we compare the space-time spectra obtained from the active and passive methods, which cover different frequency ranges. The propagation of swell present low-frequency signals, in the range of 0.1-0.7~Hz. This complements the space-time spectra obtained from active sources in the 1-35 Hz frequency band. The hydro-elastic dispersion relations fitted from active or passive methods are coherent with eachother as can be seen in Figure~\ref{fig:FK_active_passive}c, and lead to comparable results at most locations along the fiber.

%Using the propagation of these hydroelastic waves, we follow the evolution of $D$ with the curvilinear coordinate $x$ and compare our results to the obtained values of $D$ using active generation of mechanical waves. The results obtained from both passive and active methods are represented in figure \ref{fig:EhD_vs_x} (c). 

%\seb{Present evolution of $D(x)$ : three areas that coincide with the three ice regions. Coherence between instruments : DAS, geophones, ice thickness measurements. Interpretation of this evolution : jumbled ice is thicker, and younger ice thickness / Young modulus seems to decay as we get closer to the ice edge. Difference between two passive measurements, may be related to a hardening of ice ?}

\begin{figure*}[t!]
    \includegraphics[width = 1. \textwidth]{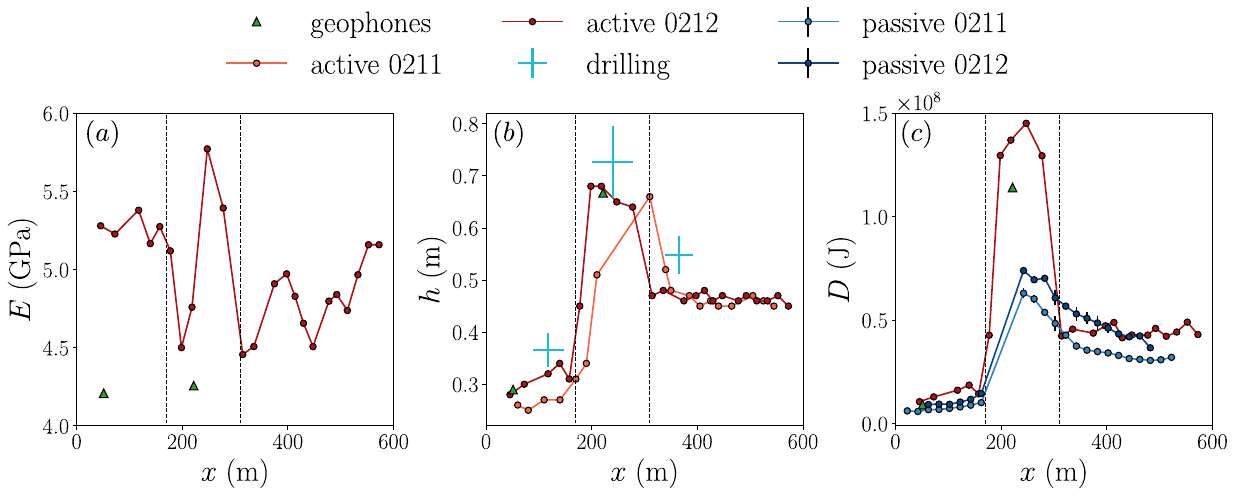}
    \caption{Variations of (a) effective Young's modulus $E$, (b) ice thickness $h$ and (c) flexural rigidity $D$ along the fiber. In each subplot, vertical dotted lines represent the frontier between the different ice regions identified in Figure \ref{fig:situation_map_bathymetry}. On 11 February, longitudinal in-plane waves created by active sources could not be measured efficiently, which explains why only the ice thickness could be inverted on this day with the active method.}
    \label{fig:EhD_vs_x}
\end{figure*}

Sea ice properties measured using both active and passive methods are presented in Figure~\ref{fig:EhD_vs_x}. Thanks to the generation of in-plane longitudinal waves in the ice plate, we measure the evolution of the Young modulus along the fiber-optic cable. The three ice regions having contrasting morphological characteristics (section~\ref{sec:method_ice}) are easily recognized. Starting from the shore side of the bay, the first ice region R$_1$ presents an effective Young modulus of about 5.2~GPa, the ice jumbled area R$_2$ shows values varying between 4.5~GPa and 5.7~GPa. The Young's modulus also fluctuates in region R$_3$ between 4.5~GPa and 5.1~GPa. We also observe a gradual increase of the ice Young's modulus as we get closer to the shore, which may be related to the ice stiffening as it gets older. The effective Young's modulus measured using the DAS system is slightly larger than the values measured by the geophones arrays (Fig.~\ref{fig:EhD_vs_x}a). The range of variations, between 4~GPa and 5.7~GPa is also in agreement with other measurements performed with geophones on first-year sea ice \citep{stein_inversion_1998, moreau_sea_2020}.

Sea ice in the Baie~du~Ha!~Ha! presents defects of finite size, hence the effective modulus may depend on the measured wavelengths, especially if wavelengths are comparable to the defects size. Using geophones, we record a larger range of frequencies, the effective Young's modulus inverted from geophones measurements may therefore account for a broad range of defect sizes, and therefore be slightly lower than values obtained using DAS.

%\steph{If there are defects of finite size, the effective modulus may depend on the wavenumber (hence on the frequency)}.

%we observe a mean effective elastic modulus $E$ of about 5.0$\pm$0.3~GPa along the fiber. \seb{Compare to bibliography} \steph{Does it make sense to talk about the average first ?} The effective Young's modulus presents variations along the fiber, especially in the ice jumbled region R$_2$.  

The evolution of the effective ice thickness along the fiber is shown in Figure~\ref{fig:EhD_vs_x}b, where the three ice regions are clearly distinct (Figure~\ref{fig:situation_map_bathymetry}). The retrieved inverted ice thicknesses are also compared with measurements from drillings. We observe a thin ice region R$_1$, with a thickness of about 0.30~m, in agreement with the effective thickness of 0.29~m measured with geophones, leading to a relative error of 3\% between the two instruments. The ice jumbled region R$_2$ appears to be about 0.67~m thick, while R$_3$ has a nearly constant thickness of 0.45~m. From the geophone array deployed in area R$_2$, we found an effective thickness of 0.66~m. 
Seismic measurements of ice thickness, using geophones or DAS, are constantly about 5~cm lower than drill-hole measurements in all three regions. We attribute this difference to be inherent to the propagation of hydro-elastic flexural waves in the plate. The bottom of the ice plate, in contact with water, appeared to be made of porous, consolidated slushy ice, which is significantly softer than upper layers of the ice plate. This soft bottom part of the ice sheet may not be able to guide elastic waves, but is stiff enough to stop the hook of the  measuring tape.
%This may explain the 5~cm shift between seismic measurements and measurements done by drilling.
The measurement techniques are complementary, where drillings  better estimate the total thickness, used for instance to survey ice volume, while seismic measurements will give access to the effective thickness governing swell propagation in ice.

Finally, both active and passive methods show a similar change in the ice flexural rigidity along the optical fiber. Region R$_1$ has a flexural rigidity ranging between $D=5.8$~MJ and $D=17$~MJ. In the jumble ice region R$_2$, $D$ increases to 67~MJ or 145~MJ according to passive or active methods, respectively. The flexural rigidity $D$ decreases over the last 300~m of the fiber, corresponding to ice region R$_3$, as both methods return values of between 30~MJ and 58~MJ. As $D \propto h^3$, the two evolve similarly along the cable.  

Compared to geophones and the passive method, we obtain higher values of $D$ in the region $R_2$ with the combination of active noise generation and DAS system. The difference between active and passive methods with the DAS system may be related to an averaging effect of the CWT computation in this specific ice region $R_2$. We observe the useful swell signal only in a relatively narrow spatial section of the fiber in this ice area. Assuming a constant effective Young's modulus in this ice area, this difference of flexural rigidity between the two methods would lead to an uncertainty on the ice thickness of 24~\%. Observations of swell propagation are therefore not sufficient to get a precise estimate of the ice thickness in highly irregular sea ice. The ice-cable mechanical coupling in such regions needs to be improved.

%These measurements obtained from the DAS system are coherent with the three ice regions we identified but also with the values of $D$ computed from the geophones linear arrays and manual ice thickness measurements. Indeed, from the effective values of Young modulus $E$, Poisson coefficient $\nu$ and ice thickness $h$, we measure over a given geophone line, we can compute a flexural rigidity $D = \frac{Eh^3}{12(1-\nu^2)}$.   

%With the linear array deployed around $x = 50 \; \mathrm{m}$, we compute a flexural rigidity $D = 9 \; \mathrm{MJ}$, coherent with values obtained from the DAS system in this first older ice region $R_1$. Around $x = 222 \; \mathrm{m}$, we measure a higher bending modulus of about $110 \; \mathrm{MJ}$, which correlates the ice thickness measurements in this jumbled area $R_2$ as well as the measurements from the DAS in the same region, at $x = 300 \; \mathrm{m}$. Considering a constant value of the ice Young modulus and Poisson coefficient over the whole fiber length, we convert each ice thickness measure to a bending modulus, as plotted in figure \ref{fig:D_vs_x}. We choose Young modulus and Poisson coefficient values of $E = 4.2 \; \mathrm{GPa}$ and $\nu = 0.27$, which correspond to the values averaged over the three geophones linear arrays. The computed bending modulus the same trend as the one obtained from the DAS system : lower values in the ice region $R_1$, higher values in the irregular region $R_2$ before a decrease in the last area $R_3$. 

\section{Discussion}
\label{sec:discussion}

Hydroelastic swell and seismic waves generated by active sources appear complementary, and enable inversion for of ice Young's modulus, thickness and flexural rigidity with good accuracy along the entire cable. We succeeded in identifying the different sea ice regions observed from the ice surface and detailed in section~\ref{sec:method_ice}. The jumbled ice region R$_2$ has higher values of effective ice thickness and flexural rigidity than the two other ice regions, associated with the piling up of several ice fragments during previous compressive events. These measurements are in good agreement with other instruments, which is promising for further studies of sea ice property variability using DAS.

The optoelectronic interrogator unit is quite energy-intensive and the absence of electric power supply in the bay constrained us to perform only hour-long acquisitions. Longer recording would allow, for instance, use of the ambient seismic noise surrounding the fiber for passive interferometry and retrieval of ice properties without the need for active sources as done by \cite{nziengui-ba_measuring_2023}. Indeed, active noise generation, and therefore human presence on ice, appears necessary to get precise estimates of sea ice mechanical properties on such timescales. This could be overcome by monitoring high frequencies icequakes, which coupled to swell seismic noise, could enable fully remote monitoring of ice properties.

Due to this short recording time as well as the harsh meteorological conditions that followed those measures, strong wind and waves conditions due to a storm, we focused our attention to understanding the spatial heterogeneities of ice mechanical properties. Yet, longer recording time, or measured frequency would allow the temporal monitoring of ice thickness, elastic and flexural moduli. A close look at Figure \ref{fig:EhD_vs_x}b and c shows a slight increase of the effective ice thickness and flexural rigidity between 11 and 12 February, which could be related to a thermodynamic growth of the ice as air temperature in Rimouski was about -12~$^{\circ}$C on the night of 11 February. Yet, swell frequencies were slightly higher on 12 February, sampling slightly smaller defects and leading to a small change of the effective flexural rigidity. Additional measurements are required to make conclusions on the temporal evolution.

Our observations also open the possibility to quantify the damping of swell by a continuous sea ice plate, as has been achieved with seafloor cables \citep{smith_observations_2023}. However, a correct calibration of the observed strain rate signal amplitude is required. This could be achieved using a few geophones placed along the cable \citep{lindsey_broadband_2020}. Yet, the heterogeneous mechanical coupling between the cable and ice could complicate this task. A regular clamping of the fiber on the ice surface seems to ensure a sufficient coupling, even though we were quite lucky that a thin layer of snow covered the fiber-optic cable and reinforced the mechanical coupling. Pouring water on top of the cable \cite{xie_ice_2024} could be a solution to ensure an homogeneous coupling along the entire fiber, but it would prolong the deployment time and would not ensure a proper recovery of the optical cable. %Despite being cheaper than seismometers, we believe that a proper recovery of multiple kilometer-long fiber made of plastic, metal and glass is preferable.

Despite being able to measure the heterogeneity in the ice thickness and Young's modulus within the bay, we believe that the use of DAS systems on sea ice in the Arctic environment is still quite complex. Extreme weather conditions, efficient power supply, mechanical coupling issues in highly irregular ice, sea ice dynamics and its possible fracture due to surface waves and other constraints make cable deployments for long recordings difficult.

\section{Conclusions}  %% \conclusions[modified heading if necessary]

We present the use of a DAS system, as a dense sensor array to measure the propagation of seismic signals in sea ice, generated by both hydro-elastic swell and hammer shots. Using plastic cases, we improved the coupling of a 600 meters-long fiber-optic with fast ice in Baie du Ha!~Ha!, located in the Bic National Park, Canada. The cable crossed three ice regions with contrasting morphological properties. We connected the optical fiber to an interrogator unit in order to measure the propagation of strain rate field within the fiber. By extracting the dispersion curves of the seismic waveforms through the different ice regions, we measure the evolution of an effective Young's and flexural moduli as well as the effective ice thickness along the entire cable length. These properties were in agreement with those obtained with geophones arrays and in-situ measurements. The Young's modulus appeared to evolve between 4.5 and 5.7~GPa, while sea ice thickness varies from 0.3~m close to the shore to 0.67~m in the most irregular ice area. This sudden change in thickness comforts the idea that the ice jumbled area was formed by ice fragments rafting, taking place during windy and high swell conditions. Further tests and technological advances are required to improve the ice-cable coupling as well as to perform long-time acquisition. However, this work suggests that the DAS system is an efficient way to map sea ice properties over large distances to better understand sea ice formation.

\section{Declaration of Competing Interests}

%The authors acknowledge that there are no conflicts of interest recorded.\footnote{The authors acknowledge that there are no conflicts of interest recorded.}

The contact author has declared that none of the authors has any competing interests.

\begin{acknowledgements}

Raw data is available at https://doi.org/10.57745/D3CJYO.\\
Codes and scripts used for data analysis are available on github: https://github.com/Turbotice/icewave.git \\
VD acknowledges support  through Schmidt Sciences, LLC, via the SASIP project (grant G-24-66154).
MMS acknowledges support from Woods Hole Oceanographic Institution's Access to the Sea program and the Office of Naval Research Award N000142412681.
DD received support from NSERC Discovery Grants \emph{Physics of seasonal sea ice} (RGPIN-2019-06563) and \emph{Wave-ice-ocean interactions and polynyas} (RGPIN-2025-05540).
This research was undertaken in part thanks to funding from the Canada First Research Excellence Fund for the Transforming Climate Action (TCA) program, and contributes to the scientific objectives of the FRQNT Strategic Cluster Québec-Océan.
LD, SP and AE acknowledge the support of the Agence Nationale de la Recherche through grant MSIM ANR-23-CE01-0020-02.
SP and AE acknowledge the support of Mairie de Paris through Emergence(s) grant 2021-DAE-100245973, of the Agence Nationale de la Recherche through grant TRANSWAVES ANR-24-CE51-3840.

\end{acknowledgements}

%\bibliography{biblio}

\end{document}